\documentclass[aps,twocolumn,pre,superscriptaddress,showpacs]{revtex4-1}

\usepackage{amsmath}
\usepackage{amssymb}
\usepackage{graphicx}

\begin{document}

\title{Role of the on-site pinning potential in establishing 
quasi-steady-state conditions of heat transport in finite quantum systems}

\author{Eduardo C. Cuansing}
\email{eduardo.cuansing@gmail.com}
\affiliation{Department of Electrical and Computer Engineering, National
University of Singapore, Singapore 117576, Republic of Singapore}
\author{Huanan Li}
\author{Jian-Sheng Wang}
\affiliation{Department of Physics and Centre for Computational Science
and Engineering, National University of Singapore, Singapore 117542,
Republic of Singapore}

\date{22 May 2012; published 24 September 2012}

\begin{abstract}

We study the transport of energy in a finite linear harmonic chain by 
solving the Heisenberg equation of motion, as well as by using 
nonequilibrium Green's functions to verify our results. The initial state 
of the system consists of two separate and finite linear chains that are
in their respective equilibriums at different temperatures. The chains are
then abruptly attached to form a composite chain. The time evolution of 
the current from just after switch-on to the transient regime and then
to later times is determined numerically. We expect the current to 
approach a steady-state value at later times. Surprisingly, this is 
possible only if a nonzero quadratic on-site pinning potential is applied 
to each particle in the chain. If there is no on-site potential a 
recurrent phenomenon appears when the time scale is longer than the 
traveling time of sound to make a round trip from the midpoint to a chain 
edge and then back. Analytic expressions for the transient and 
steady-state currents are derived to further elucidate the role of the 
on-site potential.

\end{abstract}

\pacs{05.70.Ln,44.10.+i,63.22.-m}

\maketitle

\section{Introduction}
\label{sec:intro}

In the study of thermal transport in quantum systems, the conventional
setup examined is a scattering region sandwiched between two infinite
leads acting as large heat reservoirs \cite{wang08}. The leads and the 
scattering region are initially prepared to be at their respective thermal 
equilibriums in canonical distributions and then are coupled using an 
adiabatic switch-on. The long-time steady-state current through the 
scattering region can be calculated using the Landauer formula where the 
transmission coefficient can be determined using nonequilibrium Green's 
function techniques \cite{wang08}. This approach, however, is not 
applicable to situations where the system is undergoing dynamical changes 
and is far from the steady-state regime. An example would be the transient 
behavior of a system where the coupling between the scattering region and 
the leads is not weak and is switched on abruptly \cite{cuansing10}. In 
addition, the method does not provide any information on how the steady 
state is dynamically approached from the initial configuration of the 
system. In this work we examine the time evolution of the thermal current 
from the transient regime just after an abrupt switch-on to the 
intermediate-time quasi-steady-state regime and then to the long-time 
recurrent regime, in a system consisting only of two finite leads that 
are coupled abruptly at time $t = 0$. We examine if, when, and how the 
onset of the steady-state occurs. To determine the current, we solve the 
quantum equations of motion of the linear system. In addition, a 
supplementary second method we employ to verify our results is an approach 
using nonequilibrium Green's functions that takes into account the full 
time evolution of the system \cite{cuansing10}.

\begin{figure}[h!]
\includegraphics[width=3.3in,clip]{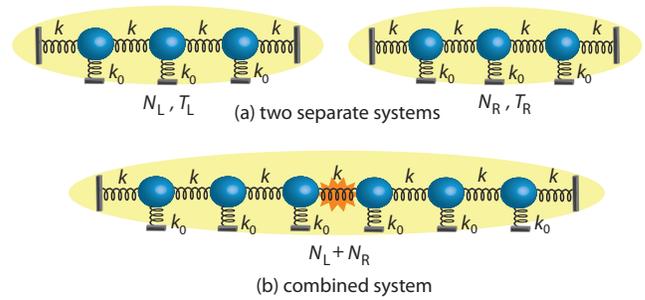}
\caption{(Color online) An illustration of how two finite independent
systems at temperature $T_{\rm L}$ and $T_{\rm R}$, containing $N_{\rm L}$ 
and $N_{\rm R}$ particles, are abruptly combined at $t = 0$ into a composite 
system containing $N_{\rm L} + N_{\rm R}$ particles. The interparticle 
spring constant is $k$ and the on-site spring constant is $k_0$.
\label{fig:schematic}}
\end{figure}

We determine the thermal current in a phonon system consisting of two
linear chains abruptly attached together at time $t = 0$. Shown in 
Fig.~\ref{fig:schematic}(a) are the two separate linear chains. The 
interparticle spring constant is $k$ and the on-site spring constant is 
$k_0$. The left and right leads contain $N_L$ and $N_R$ sites and have 
temperature $T_L$ and $T_R$, respectively. The chains are initially in 
their respective thermal equilibrium; i.e., they are initially attached 
to heat baths so that they acquire their corresponding temperatures and 
then the baths are subsequently detached. The chains therefore satisfy 
canonical distributions. They also follow fixed boundary conditions 
wherein particles at the left and right edges are attached to fixed 
walls. At time $t = 0$ the chains are abruptly coupled with spring 
constant $k$, as shown in Fig.~\ref{fig:schematic}(b). The composite chain 
also satisfies fixed boundary conditions. We then determine the energy 
current flowing through the newly formed interleads coupling. Since the 
chains are finite, we cannot use the Landauer formula approach to 
calculate the current. However, results from the Landauer formula can be 
used for comparison to the infinitely large and long-time limit results 
of the approaches we describe.

The probability of extracting work in a system that is weakly coupled to 
finite heat baths is found to follow a power law \cite{campisi09}. In this
paper, in comparison, we examine the energy current flowing between two 
finite chains whose coupling is not weak. We find that the presence of a
quadratic on-site pinning potential is necessary in establishing a 
steady-state current. Previous studies have found that the on-site pinning 
potential plays an important role in how the steady-state phonon current 
depends on the system size in disordered harmonic systems \cite{pinning}. 
In this work we further investigate the role of the on-site potential on 
the dynamics of the system and find that. without a quadratic on-site 
pinning potential, the current exhibits oscillatory behavior that does not 
disappear even for long times. Furthermore, we find that the time period 
of oscillation of the current is proportional to the sum of the length of 
the finite chains. Studies in systems consisting of classical harmonic 
oscillators and classical particles indicate that the Poincar{\'{e}} 
recurrence time, i.e., the time for a specific phase-space configuration 
of the system, or a configuration that very closely resembles it, to 
reappear increases exponentially with the number of degrees of freedom 
\cite{frisch56,zwanzig01}. Recurrence of the wave function is also found 
to occur in quantum systems with discrete energy spectrum \cite{bocchieri59} 
and systems that are periodically driven \cite{hogg82}. Our work extends 
these studies on quantum systems by examining a physical observable, i.e., 
the energy current, in a finite system to see if it displays recurrent 
behavior and to determine conditions that dictate the appearance of 
recurrence. In Sec.~\ref{sec:results} our results show that the presence 
of an on-site potential is crucial to determine whether the energy 
current exhibits a recurrent and oscillatory behavior or a behavior that 
decays and settles to a quasi-steady-state value.

In electron systems, the transport of electrons between two finite leads 
has previously been studied \cite{bushong05}. When there is a potential 
bias between the leads, it is found that a quasi-steady-state current 
with a finite lifetime appears, even when there are no dissipative effects 
like electron-electron and electron-ion interactions. In this paper, we 
supplement this electron transport study by examining the transport of 
phonons within a finite system. We determine the dynamical behavior of 
the thermal current and find that having on-site pinning potentials is 
necessary in establishing the quasi-steady-state current.

\section{The eigenmode approach}
\label{sec:eigenmode}

We model the left and right chains by the harmonic Hamiltonian
\begin{equation}
H^{\alpha} = \frac{1}{2} (p^{\alpha})^T p^{\alpha} + \frac{1}{2}
(u^{\alpha})^T K^{\alpha} u^{\alpha},~~~\alpha = {\rm L, R},
\label{eq:hamiltonian}
\end{equation}
where $u^{\alpha}$ is a column vector whose elements are the renormalized 
displacements of the sites in chain $\alpha$, $p^{\alpha}$ is the conjugate 
momentum, and $K^{\alpha}$ is the tridiagonal spring constant matrix 
consisting of $2 k + k_0$ along the diagonal and $-k$ along the two 
off-diagonals. $k$ is the nearest-neighbor spring constant and $k_0$ is 
the spring constant of the on-site pinning potential. The spring potentials
we consider in this work are all quadratic. The equations of motion of 
particles in the left chain are
\begin{equation}
\dot{u}^{\rm L} = p^{\rm L}~~~{\rm and}~~~
\ddot{u}^{\rm L} = \dot{p}^{\rm L} = -K^{\rm L} u^{\rm L} - V^{\rm LR}
u^{\rm R},
\label{eq:eqnmotion}
\end{equation}
where $V^{\rm LR}$ is the coupling matrix between the left and right chains.
Particles in the right chain satisfy a similar set of equations of motion.
The current flowing out of the left chain can be calculated from the chain's
average rate of decrease in energy,
\begin{align}
I^{\rm L} = {} & -\left< \dot{H}^{\rm L} \right>
= \left< p^{\rm L}(t)^T~V^{\rm LR}~u^{\rm R}(t) \right> \nonumber \\
= {} & \sum_{jk} \left<p_j^{\rm L}(t)~V_{jk}^{\rm LR}~u_k^{\rm R}(t)\right>,
\label{eq:Ileft}
\end{align}
where the equations of motion in Eq.~(\ref{eq:eqnmotion}) are used. The
average is taken with respect to the initial product state density matrix. 
To calculate the energy current out of the left chain, therefore, we need
to determine the dynamics of $u^{\rm R}(t)$ and $p^{\rm L}(t)$.

Consider an isolated finite linear chain consisting of $N$ sites and
satisfying fixed boundary conditions. Since it is isolated, the equations
of motion of its constituent particles are
\begin{equation}
\dot{u} = p~~~{\rm and}~~~\ddot{u} = \dot{p} = -K\,u,
\label{eq:isolate}
\end{equation}
where $K$ is the spring constant matrix of the whole chain. The solution is
\begin{equation}
\begin{split}
u(t) & = \cos(\sqrt{K} t)\,u_0 + \frac{1}{\sqrt{K}} \sin(\sqrt{K} t)
\,p_0, \\
p(t) & = -\sqrt{K} \sin(\sqrt{K} t)\,u_0 + \cos(\sqrt{K} t)\,p_0,
\label{eq:solution}
\end{split}
\end{equation}
where $u_0$ and $p_0$ are determined from the initial condition.  Since the
system is linear, quantum Heisenberg operators and classical variables 
have identical solutions. To make the solution numerically amenable, we 
construct a transformation, using eigenmodes, that diagonalizes the matrix 
$K$. Let $u^n$ be the $n$th eigenmode of $K$, i.e., 
$K\,u^n = \Omega_n^2\,u^n$, where $\Omega_n^2$ is the eigenvalue associated 
with $u^n$, i.e.,
\begin{equation}
\Omega_n^2 = 2 k \left( 1 - \cos q_n \right) + k_0,
\label{eq:eigenvalue}
\end{equation}
where $q_n = \pi n/(N+1)$ and  $n = 1, 2, \cdots, N$. Because of the fixed 
boundary conditions, the $j$th element of the eigenmode is 
$u_j^n = A_n\,\sin(q_n\,j)$, where $A_n$ can be fixed by an appropriate 
normalization. Let $S$ be a matrix consisting of the eigenmodes of $K$, 
i.e., $S = ( u^1, u^2, \ldots, u^N)$. We can then fix $A_n$ by normalizing 
$S S^T = S^T S = 1$. We get
\begin{equation}
A_n = \sqrt{\frac{2}{N+1}}.
\label{eq:An}
\end{equation}
The matrix $K$ is diagonalized by the similarity transformation
\begin{equation}
S^T K S = {\rm diag}( \Omega_1^2, \Omega_2^2, \ldots, \Omega_N^2) \equiv 
\Omega^2,
\label{eq:diagonal}
\end{equation}
where the right-hand side is a diagonal matrix consisting of elements
$\Omega_1^2$, $\Omega_2^2$, $\ldots$, $\Omega_N^2$. The coefficients
involving $K$ in Eq.~(\ref{eq:solution}) can now be calculated with the
aid of the above similarity transformation. We are then left with the
unknowns $u_0$ and $p_0$, and their correlations. To determine these 
unknowns, we write $u_0$ in normal mode coordinates, i.e., $u_0 = S Q$, 
where $Q$ contains the normal modes. In a harmonic oscillator with 
Hamiltonian $H_{ho} = \sum_j \hbar \Omega_j ( a_j^{\dagger} a_j + 1/2 )$, 
the $j$th normal mode is
\begin{equation}
\begin{split}
Q_j & = \sqrt{\frac{\hbar}{2 \Omega_j}} ( a_j + a_j^{\dagger} ), \\
P_j & = \dot{Q}_j = -i \Omega_j \sqrt{\frac{\hbar}{2 \Omega_j}}
( a_j - a_j^{\dagger} ),
\label{eq:normalmodes}
\end{split}
\end{equation}
where $a_j(t) = a_{j}(0) e^{-i \Omega_j t}$ and its complex conjugate are
the time-dependent lowering and raising ladder operators. These ladder
operators satisfy expectation values
$\left<a_j^{\dagger} a_k\right> = \delta_{jk} f_j$,
$\left<a_j a_k^{\dagger}\right> = \delta_{jk} ( f_j + 1 )$, and
$\left<a_j^{\dagger} a_k^{\dagger}\right> = \left<a_j a_k\right> = 0$,
where $f_j = (\exp(\hbar \Omega_j/k_B T) - 1)^{-1}$ is the Bose-Einstein
distribution function. Using these expectation values, we get
\begin{equation}
\begin{split}
\left<Q_j Q_k\right> & = \frac{\hbar}{2 \Omega_j} \left(2 f_j + 1\right)
\delta_{jk}, \\
\left<Q_j P_k\right> & = \frac{i \hbar}{2} \delta_{jk},  \\
\left<P_j Q_k\right> & = -\frac{i \hbar}{2} \delta_{jk}, \\
\left<P_j P_k\right> & = \Omega_j \frac{\hbar}{2} \left(2 f_j + 1\right)
\delta_{jk}.
\label{eq:QP}
\end{split}
\end{equation}
Using $S$ and the above expectation values, we can write
\begin{equation}
\begin{split}
\left<u_{0 j} u_{0 k}\right> & = \sum_m S_{jm} \frac{\hbar}{2 \Omega_m}
\left(2 f_m + 1\right) S_{km}, \\
\left<u_{0 j} p_{0 k}\right> & = \sum_m S_{jm} \frac{i \hbar}{2}
S_{km}, \\
\left<p_{0 j} u_{0 k}\right> & = -\sum_m S_{jm} \frac{i \hbar}{2}
S_{km}, \\
\left<p_{0 j} p_{0 k}\right> & = \sum_m S_{jm} \frac{\hbar}{2}
\left(2 f_m + 1\right) \Omega_m S_{km}.
\label{eq:up}
\end{split}
\end{equation}
For the composite chain in Fig.~\ref{fig:schematic}(b), we label the sites
consecutively from $1$ for the leftmost site to $N_{\rm L}+N_{\rm R}$ for
the rightmost site. Thus, the labels of the sites where the newly formed
interleads spring appears are $N_{\rm L}$ and $N_{\rm L}+1$. In
Eq.~(\ref{eq:Ileft}) therefore, the indices are $j = N_{\rm L}$ and
$k = N_{\rm L}+1$. The expectation value in Eq.~(\ref{eq:Ileft}), using the
solutions in Eq.~(\ref{eq:solution}), can then be written as
\begin{equation}
\left<p_{N_{\rm L}}^{\rm L}(t)\,u_{N_{\rm L}+1}^{\rm R}(t)\right>
= \sum_{j,k = 1}^{N_{\rm L}} \Upsilon_{jk}^{\rm L}
+ \sum_{j,k=N_{\rm L}+1}^{N_{\rm L}+N_{\rm R}} \Upsilon_{jk}^{\rm R},
\label{eq:pLuR}
\end{equation}
where
\begin{equation}
\begin{split}
\Upsilon_{jk}^{\alpha} & = C_j\,A_k \left<u_{0 j}^{\alpha} 
u_{0 k}^{\alpha}\right> + C_j\,B_k \left<u_{0 j}^{\alpha} 
p_{0 k}^{\alpha}\right> \\
& + D_j\,A_k \left<p_{0 j}^{\alpha} u_{0 k}^{\alpha}\right> 
+ D_j\,B_k \left<p_{0 j}^{\alpha} p_{0 k}^{\alpha}\right>,
\end{split}
\label{eq:Ijk}
\end{equation}
and $\alpha = {\rm L, R}$.  The $\left<up\right>$ and $\left<pu\right>$ 
terms cancel exactly and only the $\left<uu\right>$ and $\left<pp\right>$ 
correlations contribute to the currents. Note also that correlation 
between the left and right regions vanishes and each region has its own 
initial temperature $T_{\alpha}$. The coefficients are the symmetrized 
version of those in Eq.~(\ref{eq:solution}):
\begin{equation}
\begin{split}
A_k & = \sum_{m=1}^N S_{N_{\rm L}+1,m} \cos\left(\Omega_{m} t\right)
S_{km}, \\
B_k & = \sum_{m=1}^N S_{N_{\rm L}+1,m} \frac{\sin\left(\Omega_{m}
t\right)}{\Omega_{m}} S_{km}, \\
C_j & = -\sum_{m=1}^N S_{N_{\rm L},m} \Omega_{m} \sin\left(\Omega_{m}
t\right) S_{jm}, \\
D_j & = \sum_{m=1}^N S_{N_{\rm L},m} \cos\left(\Omega_{m} t\right) S_{jm},
\label{eq:ABCD}
\end{split}
\end{equation}
where $N = N_{\rm L} + N_{\rm R}$. The current can therefore be calculated
using Eqs.~(\ref{eq:Ileft}), (\ref{eq:pLuR}), (\ref{eq:Ijk}), (\ref{eq:up}), 
and (\ref{eq:ABCD}). In addition, Eq.~(\ref{eq:pLuR}) can be further 
simplified into two terms. One of the terms is proportional to 
$(f_{\rm L} - f_{\rm R})$. This term leads to the steady-state value in the 
limit $N \rightarrow \infty$. The other term is proportional to 
$(f_{\rm L} + f_{\rm R} + 1/2)$ and produces the transient current. See the 
Appendix.

The expression for $I^{\rm L}(t)$ in Eq.~(\ref{eq:Ileft}) is the current
flowing through the rightmost site of the left lead, i.e., at the site
labeled $N_{\rm L}$. At any site $n$ in the composite lead, a general 
expression for the current flowing through it can be written as
\begin{equation}
\begin{split}
I^n(t) & = V_{nm}\, \left< p_n(t)\, u_m(t)  \right> \\
& = V_{nm} \left\{ \sum_{j,k=1}^{N_{\rm L}} \Upsilon_{nm,jk}^{\rm L} 
+ \sum_{j,k=N_{\rm L}+1}^{N_{\rm L}+N_{\rm R}} \Upsilon_{nm,jk}^{\rm R} 
\right\}
\label{eq:Iany}
\end{split}
\end{equation}
where $V_{nm}$ is the coupling between sites $n$ and $m=n+1$ and
\begin{equation}
\begin{split}
\Upsilon_{nm,jk}^{\alpha} & = C_{nj} A_{mk} \left< u_{0j}^{\alpha} 
u_{0k}^{\alpha} \right> + C_{nj} B_{mk} \left< u_{0j}^{\alpha} 
p_{0k}^{\alpha} \right> \\
& + D_{nj} A_{mk} \left< p_{0j}^{\alpha} u_{0k}^{\alpha} \right>
+ D_{nj} B_{mk} \left< p_{0j}^{\alpha} p_{0k}^{\alpha} \right>,
\label{eq:Inmjk}
\end{split}
\end{equation}
where $\alpha = {\rm L}, {\rm R}$. Equation (\ref{eq:Inmjk}) is in the same 
form as Eq.~(\ref{eq:Ijk}) but generalized to include the current flowing 
through any site $n$ in the composite chain. Similarly, the coefficients 
$A_{mk}$, $B_{mk}$, $C_{nj}$, and $D_{nj}$ are the generalized forms of 
those in Eq.~(\ref{eq:ABCD}).

\section{Nonequilibrium Green's functions approach}
\label{sec:negf}

An alternative approach is to use nonequilibrium Green's functions (NEGF)
techniques to calculate the time-dependent energy current~\cite{cuansing10}.
The retarded Green's function for a finite collection of harmonic 
oscillators in a chain in equilibrium can be written as \cite{jiang09}
\begin{equation}
g^r(t) = -S~\theta(t)\; \frac{\sin\left(\Omega t\right)}{\Omega}\; S^T,
\label{eq:gr}
\end{equation}
where $S$ is the matrix in Eq.~(\ref{eq:diagonal}) and $\Omega$ is the 
diagonal matrix of the square root of the eigenvalues in 
Eq.~(\ref{eq:eigenvalue}). The advanced, lesser, and greater equilibrium 
Green's functions can then be determined from the above retarded Green's 
function. From these equilibrium Green's functions, the time-dependent 
nonequilibrium Green's functions and the energy current can be calculated 
following the procedure described in Ref.~\cite{cuansing10}. With the 
eigenmode approach in Sec.~\ref{sec:eigenmode} and the NEGF approach 
discussed in this section, we can therefore calculate the energy current in 
two different and independent ways. We find that both methods produce the 
exact same results, up to double-precision accuracy.

\section{Numerical results and discussion}
\label{sec:results}

The energy current can be calculated using either the eigenmode approach
discussed in Sec.~\ref{sec:eigenmode} or the NEGF approach described in
Sec.~\ref{sec:negf}. However, the NEGF calculation is computationally
intensive because of the presence of several multiple integrals whose
numerical convergence must be carefully determined. In contrast, 
calculations in the eigenmode approach involve, as the most complicated 
part, fast and straightforward matrix manipulations. After verifying that 
our results are the same for several sets of data from both approaches, we 
proceed and acquire most of our data using the eigenmode approach.

\begin{figure}[h!]
\includegraphics[width=3.2in,clip]{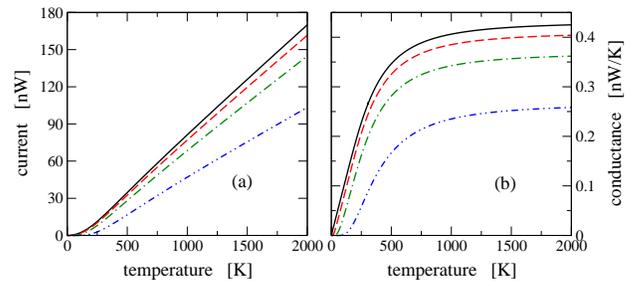}
\caption{(Color online) Plots of (a) the steady-state current and (b)
the conductance as functions of the average temperature between the two
leads and with a temperature bias of $\alpha = 0.1$. The values of the 
on-site spring constant are $k_0 = 0$ (black line), 
$k_0 = 0.01~{\rm eV/(\AA^2 u)}$ (red dash line),
$k_0 = 0.1~{\rm eV/(\AA^2 u)}$ (green dash-dot line), and
$k_0 = 1~{\rm eV/(\AA^2 u)}$ (blue dash-double dot line). The 
interparticle spring constant is $k = 1~{\rm eV/(\AA^2 u)}$.
\label{fig:steadystate}}
\end{figure}

We first determine the steady-state heat current and conductance of an 
infinite linear chain using the Landauer formula with unit transmission for 
frequencies within the phonon band \cite{wang07}. The value of the 
nearest-neighbor spring constant we use is $k = 1~{\rm eV/\AA^2 u}$. We 
examine the steady-state current for on-site pinning potentials 
$k_0 = 0, 0.01, 0.1$, and $1~{\rm eV/\AA^2 u}$. Shown in
Fig.~\ref{fig:steadystate} are the plots of the steady-state thermal current
and conductance as a function of the temperature. We use these results for 
comparison to the quasi-steady-state values arising in finite leads.

\begin{figure}[h!]
\includegraphics[width=3.2in,clip]{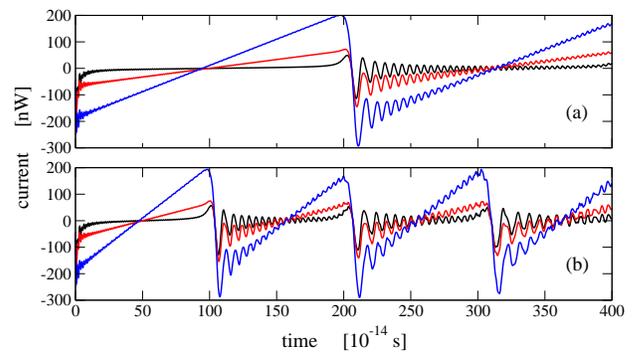}
\caption{(Color online) Plots of the current as a function of time when
the left and right chains are finite with lengths
${\rm (a)}~N_{\rm L} = N_{\rm R} = 100$ and
${\rm (b)}~N_{\rm L} = N_{\rm R} = 50$. The average temperatures between
the leads are $T = 10~{\rm K}$ (black line), $T = 100~{\rm K}$ [red (gray) 
line], and $T = 300~{\rm K}$ [blue (dark gray) line]. The temperature bias 
between the leads is $\alpha = 0.1$. There is no on-site potential, i.e., 
$k_0 = 0$.
\label{fig:k0=0000}}
\end{figure}

\begin{figure}[h!]
\includegraphics[width=3.2in,clip]{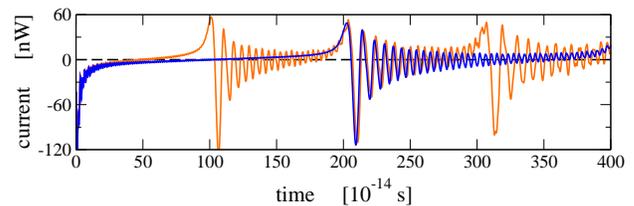}
\caption{(Color online) A close up view of the current when the average
temperature of the leads is $T = 10~{\rm K}$. The darker (blue) line is
when $N_{\rm L} = N_{\rm R} = 100$ while the lighter (orange) line is when
$N_{\rm L} = N_{\rm R} = 50$. The dashed line is the value of the
steady-state current calculated using the Landauer formula.
\label{fig:T=10K}}
\end{figure}

The dynamical energy currents are determined for various lengths and
initial temperatures of the finite leads and the strengths of the on-site 
spring potential. Shown in Fig.~\ref{fig:k0=0000} are plots of the current 
flowing out of the left lead when there is no on-site spring potential. The 
length of the leads are $N_{\rm L} = N_{\rm R} = 100$ in 
Fig.~\ref{fig:k0=0000}(a) and $N_{\rm L} = N_{\rm R} = 50$ in 
Fig.~\ref{fig:k0=0000}(b). The left lead initially has temperature 
$T_{\rm L} = (1+\alpha) T$, where $\alpha = 0.1$, while the right lead has 
initial temperature $T_{\rm R} = (1-\alpha) T$. The plots in 
Fig.~\ref{fig:k0=0000} correspond to $T = 10~{\rm K}$, $100~{\rm K}$, and 
$300~{\rm K}$. The leads are attached at time $t = 0$. Just after the
switch-on, the current initially shoots down towards negative values. Since
we are calculating the current flowing out of the left lead, a negative
current value implies that energy is flowing into the left lead. For the
right lead, we also find the same initial negative current values, i.e., 
energy is also flowing into the right lead. This result is the same as that 
found in systems with semi-infinite leads \cite{cuansing10}. The reason is 
because an external input energy is required to make the interleads coupling
between the two previously unattached leads. After the leads are connected,
the extra energy that is applied to make the connection then flows into the
leads, thus initially producing negative current values just after the 
switch-on. This phenomenon also occurs for the leads-center system and is 
explained by a small $t$ expansion \cite{bijay12}.

On longer time scales, the current rises almost linearly and then suddenly 
drops. The overall behavior is roughly periodic with a period proportional
to the full length of the chain. A ringing current can also be observed 
in Fig.~\ref{fig:k0=0000}. This ringing current also appears in systems 
containing infinite leads \cite{cuansing10}. Such a behavior has also been 
observed previously in NEGF calculations in electronic transport 
\cite{electronic}. In Fig.~\ref{fig:k0=0000}, notice that the current does 
not stabilize to a quasi-steady-state value, even when the leads have the 
low average temperature $T = 10~{\rm K}$, as shown in Fig.~\ref{fig:T=10K}.

\begin{figure}[h!]
\includegraphics[width=3.2in,clip]{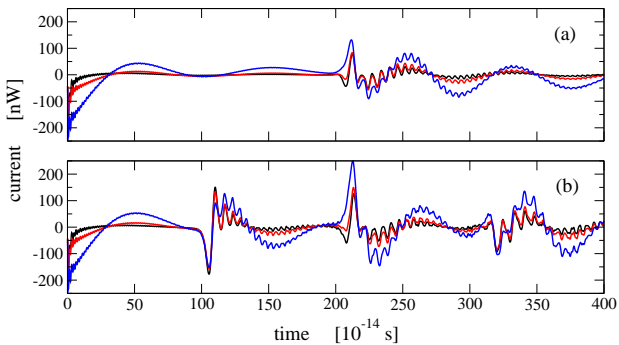}
\caption{(Color online) Plots of the current as a function of time when
the on-site spring constant is $k_0 = 0.001~{\rm eV/\AA^2 u}$. The
left and right chains have lengths ${\rm (a)}~N_{\rm L} = N_{\rm R} = 100$
and ${\rm (b)}~N_{\rm L} = N_{\rm R} = 50$. The average temperatures
between the chains are $T = 10~{\rm K}$ (black line), $T = 100~{\rm K}$
[red (gray) line], and $T = 300~{\rm K}$ [blue (dark gray) line]. The 
temperature offset is $\alpha = 10\%$.
\label{fig:k0=0001}}
\end{figure}

We next examine what happens to the current in the presence of a small
on-site quadratic potential with on-site spring constant 
$k_0 = 0.001~{\rm eV/\AA^2 u}$. Shown in Fig.~\ref{fig:k0=0001} are plots 
of the current as it evolves in time. Compared to the current shown in 
Fig.~\ref{fig:k0=0000}, although there is still the initial overshoot to a
negative value, the current appears to approach a quasi-steady-state
value. However, the value of the on-site potential is still not enough to 
suppress the large current oscillations, especially when the temperature 
is high.

\begin{figure}[h!]
\includegraphics[width=3.2in,clip]{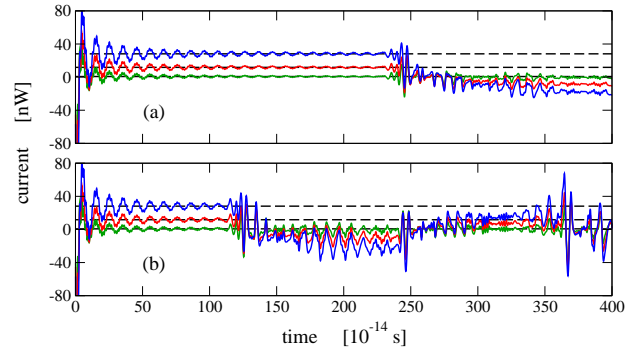}
\caption{(Color online) Plots of the current when the on-site spring
constant is $k_0 = 0.1~{\rm eV/\AA^2 u}$. The left and right chains have
lengths ${\rm (a)}~N_{\rm L} = N_{\rm R} = 100$ and
${\rm (b)}~N_{\rm L} = N_{\rm R} = 50$. The average temperatures between
the chains are $T = 100~{\rm K}$ [green (lowest gray) line], 
$T = 300~{\rm K}$ [red (middle gray) line], and $T = 500~{\rm K}$ 
[blue (upper gray) line]. The temperature offset is $\alpha = 10\%$. The 
dash lines are the values of the steady-state current, corresponding to 
$T = 100~{\rm K}$, $300~{\rm K}$, and $500~{\rm K}$, calculated 
independently from the Landauer formula.
\label{fig:k0=0100}}
\end{figure}

Based on our results for the small on-site $k_0$, we expect the current to
show a more prominent approach to a quasi-steady-state value when we increase
the value of the on-site $k_0$ further. Shown in Fig.~\ref{fig:k0=0100} are
plots of the current when $k_0 = 0.1~{\rm eV/\AA^2 u}$, i.e., at $10\%$ of
the value of $k$. Also shown in the figure are dashed lines representing the
steady-state current calculated using the Landauer formula. We now see that 
the current approaches a quasi-steady-state value and that this 
quasi-steady-state lasts longer for longer leads. Note, however, that the 
quasi-steady-state lasts for more than $2 t_m$ ($t_m$ is defined to 
be the time when sound waves travel the left or right chain). After time 
$2 t_m$, the waves or disturbances that have been reflected back at the 
hard walls at the edges of the leads have returned back to the interleads 
coupling and interfere with the other waves there. This results in the 
current beginning to oscillate wildly.

\begin{figure}[h!]
\includegraphics[width=3.2in,clip]{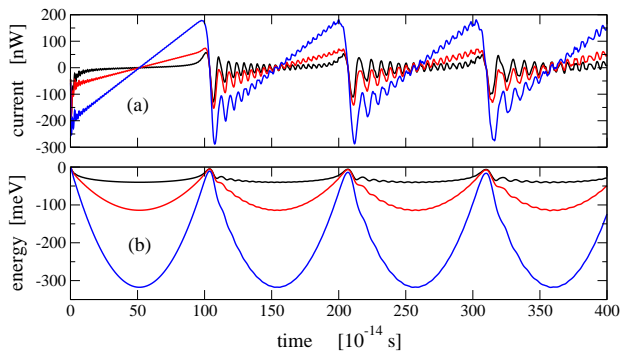}
\caption{(Color online) Plots of (a) the current and (b) the energy as
functions of time when $k_0 = 0$. The left and right chains have the same 
temperature, i.e., $\alpha = 0$, with temperatures $T = 10~{\rm K}$ (black 
line), $T = 100~{\rm K}$ [red (gray) line], and $T = 300~{\rm K}$ [blue
(dark gray) line]. The chains have length $N_{\rm L} = N_{\rm R} = 50$.
\label{fig:a=00_k0=0000}}
\end{figure}

We next set the leads to have the same temperature and, therefore, according
to the Landauer formula, we should not expect to have current flowing in the 
long-time steady-state limit. We do want to know if we get a nonzero 
transient current and, if so, how would it behave even when there is no 
temperature bias. Shown in Fig.~\ref{fig:a=00_k0=0000} are plots of the 
current flowing out of the left lead when the leads have the same 
temperature and there is no on-site potential. Also shown are plots of the 
total energy that has flowed out of the left lead. The total energy at time 
$t$ is calculated by taking the area under the curve for the current up to 
time $t$. Since the leads are indistinguishable, the current plots for the 
left and right leads are exactly the same.

Energy is added to the system when the leads are initially attached. This
energy flows into both the left and the right leads. So even when there is
no temperature bias, because of the externally added energy, a thermal
current appears in the system. However, because there is no on-site pinning 
potential, the current does not settle into a quasi-steady-state value. 
The phonons are elastically bouncing back and forth between the fixed 
walls at the edges of the two leads.

\begin{figure}[h!]
\includegraphics[width=3.2in,clip]{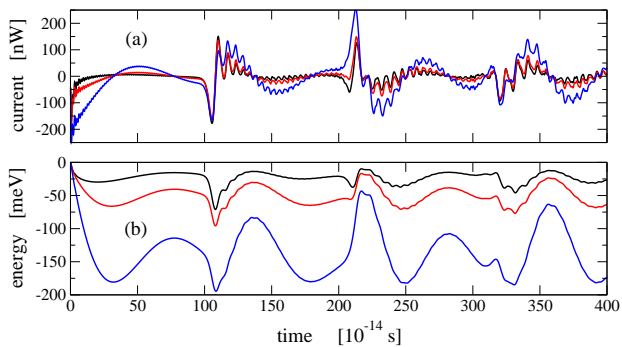}
\caption{(Color online) Plots of (a) the current and (b) the energy as
functions of time when the on-site spring constant is
$k_0 = 0.001~{\rm eV/\AA^2 u}$. The left and right chains have the
same temperature of values $T = 10~{\rm K}$ (black line),
$T = 100~{\rm K}$ [red (gray) line], and $T = 300~{\rm K}$ [blue (dark gray)
line]. The chains have length $N_{\rm L} = N_{\rm R} = 50$.
\label{fig:a=00_k0=0001}}
\end{figure}

Shown in Fig.~\ref{fig:a=00_k0=0001} are plots of the current and total
energy when a small on-site potential with $k_0 = 0.001~{\rm eV/\AA^2 u}$
is present in the system. The presence of the on-site pinning potential 
suppresses the current to a quasi-steady-state value. The total energy 
shown in Fig.~\ref{fig:a=00_k0=0001}(b) also displays a tendency to settle 
down to a quasi-steady-state value.

\begin{figure}[h!]
\includegraphics[width=3.2in,clip]{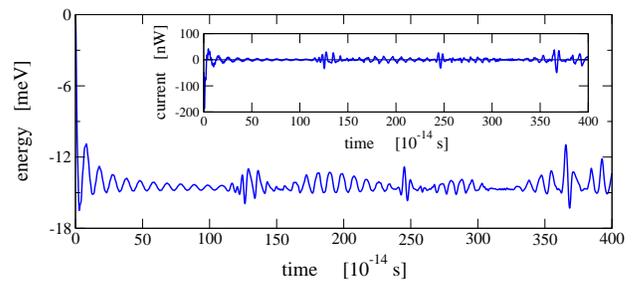}
\caption{(Color online) Plot of the energy as a function of time. The
inset shows the plot of the current as a function of time. The left and
right chains have the same temperature $T = 300~{\rm K}$ and the on-site
spring constant is $k_0 = 0.1~{\rm eV/\AA^2 u}$. The chains have
length $N_{\rm L} = N_{\rm R} = 50$.
\label{fig:a=00_k0=0100}}
\end{figure}

Shown in Fig.~\ref{fig:a=00_k0=0100} are plots of the current and total
energy when the on-site potential has spring constant
$k_0 = 0.1~{\rm eV/(\AA^2 u)}$. The value is large enough for the current 
to settle into a quasi-steady-state value. However, since the leads are 
finite, the system only has limited time before the reflected phonons 
begin to arrive and interfere with the other phonons moving through the 
interleads coupling.

\begin{figure}[h!]
\includegraphics[width=3.2in,clip]{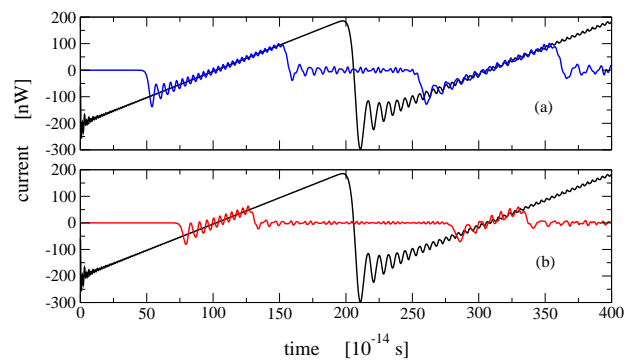}
\caption{(Color online) Plots of the current at various locations on the 
right lead. The length of the left and right leads are 
$N_{\rm L} = N_{\rm R} = 100$. The darker (black) lines in both panels (a) 
and (b) are the current at $N_{\rm L}$. The current at $n_1=N_{\rm L}+50$ 
[blue (dark gray) line in panel (a)] and $n_2=N_{\rm L}+75$ [red (gray) 
line in panel (b)] are also shown. The on-site spring constant is 
$k_0 = 0$ and the temperature $T = 300~{\rm K}$.
\label{fig:elsewhere_k0=0}}
\end{figure}

\begin{figure}[h!]
\includegraphics[width=3.2in,clip]{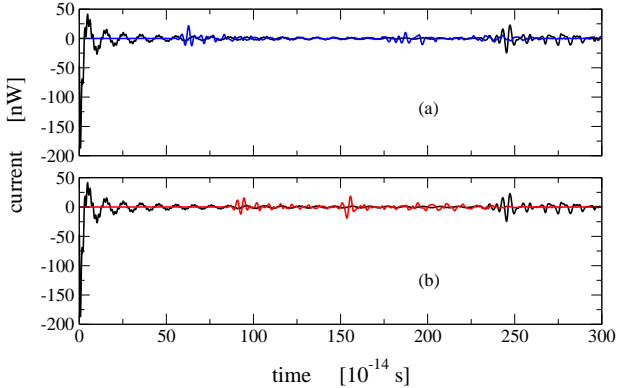}
\caption{(Color online) Plots of the current at different locations on the
right lead when $N_{\rm L} = N_{\rm R} = 100$, $T = 300~{\rm K}$, and on-site
spring constant $k_0 = 0.1~{\rm eV/\AA^2 u}$. The darker (black) lines in 
both panels (a) and (b) are the current at $N_{\rm L}$. The current at $n_1$ 
[blue (dark gray) line in panel (a)] and $n_2$ [red (gray) line in panel 
(b)] are also shown.
\label{fig:elsewhere_k0=0.1}}
\end{figure}

We can also calculate the current at sites away from the midpoint of the
composite system where the sudden attachments occurred using 
Eq.~(\ref{eq:Iany}). For a composite lead containing $N_{\rm L} = 100$ and 
$N_{\rm R} = 100$ sites, we determine the current at sites 
$n_1 = N_{\rm L} + 50$ and $n_2 = N_{\rm L} + 75$. Shown in 
Figs.~\ref{fig:elsewhere_k0=0} and \ref{fig:elsewhere_k0=0.1} are plots of 
the current at those locations. In Fig.~\ref{fig:elsewhere_k0=0} the spring 
constant is $1~{\rm eV/(\AA^2 u)}$ and there is no on-site potential. 
Because $n_1$ and $n_2$ are away from the midpoint, the current is going to 
take some time, depending on how fast phonons travel in the chain, to 
reach them. Defining a time unit $[{\rm t}] \equiv 10^{-14}\, {\rm s}$, it 
takes $50\, [{\rm t}]$ for the current to reach $n_1$ and 
$75\, [{\rm t}]$ to reach $n_2$, therefore implying a speed of $1$ site 
per $[{\rm t}]$.

In Fig.~\ref{fig:elsewhere_k0=0.1} a quadratic on-site potential with spring 
constant $k_0 = 0.1~{\rm eV/(\AA^2 u)}$ is added to the system. The two 
leads have the same temperature $T = 300~{\rm K}$. For the midpoint where 
the sudden connection at $t = 0$ occurs, upon connection the current 
immediately drops and then oscillates and decays until it reaches a 
quasi-steady-state value which, for this case, is zero. Away from the 
midpoint, the current takes some time to reach the site and so initially we 
see a flat line at zero. Since $k_0$ is not zero, the speed of the current 
is not exactly $1$ site per $[{\rm t}]$ as we found in 
Fig.~\ref{fig:elsewhere_k0=0}. 

\section{Summary and conclusion}
\label{sec:summary}

We examine the energy current in a quantum system consisting of two finite
chains that are abruptly attached at time $t = 0$. We calculate the current 
using an eigenmode approach that makes use of a similarity transformation
extending the normal modes of a single harmonic oscillator into a 
many-particle, but finite, chain of harmonic oscillators. In addition, we
use a nonequilibrium Green's function approach to verify some of the results 
from our normal mode calculations. We find that the presence of a
quadratic on-site pinning potential is crucial in establishing a 
quasi-steady-state current. In a finite system where there is no 
such on-site potential, the current does not settle to a quasi-steady-state 
value, even when the two original chains have the same temperature at a 
value as low as $10~{\rm K}$. The phonons would simply bounce back and 
forth elastically between the two fixed walls at the edges of the chains. 
In the presence of the on-site pinning potential, we find a tendency for 
the current to establish a quasi-steady state (also, see the Appendix). 
This crucial role of the on-site pinning potential in establishing a 
steady-state current should also be present even when the leads are 
semi-infinite. Computationally, the presence of the quadratic on-site 
pinning potential renders well-behaved Green's functions. The form of the
time evolution of the current depends on the interplay between the 
strength of the potentials, the chain length, and the initial chain 
temperatures. A quasi-steady-state current is established earlier when 
the on-site pinning potential is stronger. Furthermore, when the 
quasi-steady-state current is established, its value turns out to be the 
same as the steady-state value calculated using the Landauer formula. 
Although in the standard modeling of bosonic heat baths infinite systems 
are usually employed, we show here that integrable finite systems also 
behave like an infinite system but only within short time scales 
proportional to the system size $N$ and provided that a small on-site 
pinning potential is present. At time scales longer than $N$, we have 
recurrent and quasi-periodic behavior.  It is interesting to investigate 
what happens if the chains are nonlinear; however, the eigenmode expansion 
technique discussed in this paper is unable to handle nonlinear systems.

\begin{acknowledgments}
We thank Peter H{\"{a}}nggi, Jos{\'{e}} Garc{\'{i}}a-Palacios, Lifa Zhang, 
Jin-Wu Jiang, Meng Lee Leek, Xiaoxi Ni, Bijay Agarwalla, and Juzar Thingna 
for insightful discussions. J.S.W. acknowledges support from a URC 
research grant (Grant No. R-144-000-257-112).
\end{acknowledgments}

\appendix
\section{Recovering the Landauer formula}
\label{sec:appendix}
In this appendix, we simplify the formula for the current based on 
Eqs.~(\ref{eq:pLuR}) to (\ref{eq:ABCD}) and derive the Landauer formula in 
the large-size limit. We first introduce a new notation and then group 
the terms into two parts. The first part involves terms that are proportional 
to the difference between the Bose distributions of the left and right leads.
The rest of the terms constitutes the other part and are interpreted to be 
the transient contribution to the current. We assume that each chain has 
length $N$ and, therefore, that the total size of the composite chain is 
$2N$. Define 
\begin{equation}
\begin{split}
\left<k|\right.{\tilde{k}}\left.\right>^{\rm L} & \equiv \sum_{n_1=1}^{N}
\left<k|n_{1}\right>\! \left<n_{1}\right|{\tilde{k}}\left.\right>, \\
\left<k|\right.{\tilde{k}}\left.\right>^{\rm R} & \equiv \sum_{n_1=N+1}^{2N}
\left<k|n_{1}\right>\! \left<n_{1}-N\right|{\tilde{k}}\left.\right>, 
\label{eq:akk}
\end{split}
\end{equation}
where $\left<n\right.|{\tilde{k}}\left.\right> =
\left<\right.{\tilde{k}}|\left.n\right> \equiv
\sqrt{\frac{2}{N+1}}\, \sin({\tilde{k}}n)$,
${\tilde{k}} = \frac{\pi {\tilde{j}}}{N+1}$,
$\left<k|n\right> = \left<n|k\right> \equiv
\sqrt{\frac{2}{2N+1}} \sin(kn)$, and $k = \frac{\pi j}{2N+1}$.
Notice that $\left<k\right.|{\tilde{k}}\left.\right>^{\rm R} =
\left<k|\right.{\tilde{k}}\left.\right>^{\rm L} \left(-1\right)^{j+{\tilde{j}}}$
and $\left<\right.{\tilde{k}}|\left.k_2\right>^{\rm L}\!
\left<k_2\right.|{N+1}\left.\right> = 
\left<\right.{\tilde{k}}|\left.k_2\right>^{\rm L}\!
\left<k_2\right.|N\left.\right> \left(-1\right)^{j_{2}+1}$. The expression 
for the current $I^{\rm L}(t)$ can be separated into two parts, 
$I^{\rm L}_{{\rm trans}}(t)$ and $I^{\rm L}_{{\rm stdy}}(t)$, where the 
transient contribution is
\begin{equation}
\begin{split}
I^{\rm L}_{{\rm trans}}(t) & = \hbar\omega_{0}^{2}\, \sum_{\tilde{k}}
\left(f_{\tilde{k}}^{\rm L} + f_{\tilde{k}}^{\rm R} + 1\right)\\
& \times \sum_{k_1\, {\rm odd}}\left(\frac{\omega_{k_1}}{\omega_{\tilde{k}}} -
\frac{\omega_{\tilde{k}}}{\omega_{k_1}}\right) \sin\left(\omega_{k1}t\right)\, 
\left<N|k_1\right>\! \left<k_1\right|{\tilde{k}}\left.\right>^{\rm L}\\
& \times \sum_{k_2\, {\rm odd}}\cos\left(\omega_{k_2}t\right)\, 
\left<\right.{\tilde{k}}|\left.k_2\right>^{\rm L}\! \left<k_2|N\right>\\
-\, \hbar\omega_{0}^2 & \, \sum_{\tilde{k}}\left(f_{\tilde{k}}^{\rm L} + 
f_{\tilde{k}}^{\rm R} + 1\right)\\
& \times \sum_{k_1\, {\rm even}}\left(\frac{\omega_{k_1}}{\omega_{\tilde{k}}}
- \frac{\omega_{\tilde{k}}}{\omega_{k_1}}\right)
\sin\left(\omega_{k_1}t\right)\, \left<N\right|k_1\left.\right>\!  
\left<k_1\right|{\tilde{k}}\left.\right>^{\rm L}\\
& \times \sum_{k_2\, {\rm even}}\cos\left(\omega_{k_2}t\right)
\left<\right.{\tilde{k}}|k_2\left.\right>^{\rm L}\! 
\left<k_2|N\right>.
\label{eq:Itrans}
\end{split}
\end{equation}
and the steady-state contribution is
\begin{equation}
\begin{split}
I^{\rm L}_{{\rm stdy}}(t) & = \hbar\omega_{0}^{2}\, 
\sum_{\tilde{k}}\left(f_{\tilde{k}}^{L}-f_{\tilde{k}}^{R}\right)\\
& \times \sum_{k_1\, {\rm even}}\left(\frac{\omega_{k_1}}{\omega_{\tilde{k}}}
+ \frac{\omega_{\tilde{k}}}{\omega_{k_1}}\right)
\sin\left(\omega_{k_1}t\right)\, \left<N|k_1\right>\! 
\left<k_1\right|{\tilde{k}}\left.\right>^{\rm L}\\
& \times \sum_{k_2\, {\rm odd}}\cos\left(\omega_{k_2}t\right)
\left<\right.{\tilde{k}}|k_2\left.\right>^{\rm L}\! \left<k_2|N\right> \\
-\, \hbar\omega_{0}^2 & \, \sum_{\tilde{k}}\left(f_{\tilde{k}}^{L}
- f_{\tilde{k}}^{\rm R}\right)\\
& \times \sum_{k_1\, {\rm even}}\cos\left(\omega_{k_1}t\right)\, 
\left<N|k_1\right>\! \left<k_1\right|{\tilde{k}}\left.\right>^{\rm L}\\
& \times \sum_{k_2\, {\rm odd}}\left(\frac{\omega_{\tilde{k}}}{\omega_{k_2}}
+ \frac{\omega_{k_2}}{\omega_{\tilde{k}}}\right)
\sin\left(\omega_{k_2}t\right)\, 
\left<\right.{\tilde{k}}|\left.k_2\right>^{\rm L}\! \left<k_2|N\right>.
\label{eq:Istdy}
\end{split}
\end{equation}
In the above, the summation for $\tilde{k}$ extends over all 
$\frac{\pi \tilde{j}}{N+1}$ for ${\tilde{j}} = 1,...,N$, the summation 
involving ``$k_1$ even'' is on 
$k_1\in\left\{\frac{\pi j_1}{2N+1}\right\}_{j_1=1}^{2N}$, $j_1$ is even, 
and the summation involving ``$k_2$ odd'' is on 
$k_2\in\left\{\frac{\pi j_2}{2N+1}\right\} _{j_2=1}^{2N}$, $j_2$ is odd, 
and so on. The dispersion relation satisfied is
$\omega_{q} = \sqrt{2 k\left(1-\cos q\right)+k_{0}}$ where
$\omega_0^2 = k$,  $\omega_1^2 = k_0$,
$f_{\tilde{k}}^{\alpha} = 1/(e^{\beta_{\alpha}\hbar\omega_{\tilde{k}}}-1)$, 
and $\alpha = {\rm L,R}$. The sum in Eq.~(\ref{eq:akk}) can be carried out
analytically, resulting in
\begin{equation}
\begin{split}
\left<N|k_1\right>\! \left<k_1\right|{\tilde{k}}\left.\right>^{\rm L}
& = \frac{-1}{(2N+1)\sqrt{2 (N+1)}}\\
& \times \sin({\tilde{k}}N)\, 
\frac{\cos k_1 - \left(-1\right)^{j_1}}{\cos {\tilde{k}} - \cos k_1}.
\label{eq:Nktk}
\end{split}
\end{equation}
A similar expression can be derived for 
$\left<\right.{\tilde{k}}|\left.k_2\right>^{\rm L}\! \left<k_2|N\right>$. 
Using Eqs.~(\ref{eq:Itrans}) to (\ref{eq:Nktk}), the current can now be 
calculated in computer time proportional to ${\mathit O}(N^2)$. This is 
in contrast to NEGF calculations which go as ${\mathit O}(N^3)$ in 
computational complexity.

All of the expressions derived at this point are exact. We now make an
approximation in order to extend our calculations for the steady-state
contribution to large $N$ and eventually arrive at the Landauer 
formula. Notice that in Eq.~(\ref{eq:Nktk}) and the corresponding 
expression for $k_2$ that the terms involving 
$k_1\approx\tilde{k}\approx k_2$ would dominate the summation, especially 
when $N$ approaches infinity. Consequently,
\begin{equation}
\begin{split}
I^{\rm L}_{{\rm stdy}}(t) \approx & \, \hbar\omega_{0}^2\, 
\frac{1}{N+1} \sum_{\tilde{k}}\left(f_{\tilde{k}}^{\rm L} 
- f_{\tilde{k}}^{\rm R}\right) \sin^{4}\tilde{k}\\
& \times \left\{\frac{1}{2N+1}\, \frac{1}{2}
\sum_{k_2}\frac{\sin\left(\omega_{\tilde{k}} - \omega_{k_2}\right)t}
{\cos\tilde{k}-\cos k_2}\right\}\\
& \times \left\{\frac{1}{2N+1} \left\{ \sum_{k_1=e}
\frac{-1}{\cos\tilde{k} - \cos k_1}\right.\right.\\
& \;\;\; + \left.\left.\sum_{k_1\, {\rm odd}}
\frac{1}{\cos\tilde{k}-\cos k_1}\right\} \right\}.
\end{split}
\end{equation}
Let $N\rightarrow\infty$ and then followed by $t\rightarrow\infty$, we get
\begin{equation}
\frac{1}{2N+1}\, \frac{1}{2} \sum_{k_2}
\frac{\sin\left(\omega_{\tilde{k}} - \omega_{k_2}\right)t}
{\cos\tilde{k}-\cos k_2} \approx - \frac{1}{2\sin \tilde{k}}.
\end{equation}
Furthermore, we have
\begin{equation}
\begin{split}
\lim_{N\rightarrow\infty} \frac{1}{2N+1}
& \left\{\sum_{k_1\, {\rm even}}\frac{-1}{\cos{\tilde{k}}-\cos k_1}\right.\\
& + \left.\sum_{k_1\, {\rm odd}}\frac{1}{\cos\tilde{k}-\cos k_1}\right\} 
= - \frac{1}{\sin^2{\tilde{k}}}
\end{split}
\end{equation}
for some $\tilde{k}\in\left\{\frac{\pi\tilde{j}}{N+1}\right\} _{\tilde{j}=1}^{N}$.
We then recover the Landauer formula
\begin{equation}
\begin{split}
I^{\rm L}_{{\rm stdy}} & = \frac{1}{2}\, \hbar\omega_0^2\, 
\frac{1}{N+1} \sum_{\tilde{k}} \left(f_{\tilde{k}}^{\rm L}
- f_{\tilde{k}}^{\rm R}\right) \sin\tilde{k}\\
& =  \frac{1}{2 \pi}\int_{\omega_1}^{\sqrt{4\omega_0^2+\omega_1^2}} d\omega\, 
\hbar\omega \left(f^{\rm L}(\omega) - f^{\rm R}(\omega)\right).
\end{split}
\end{equation}
Note in the above that the discrete summation over wave vector $\tilde{k}$ is 
converted into a continuous integration over the angular frequency $\omega$. 
Furthermore, we want to emphasize that although the on-site constant $k_0$
appears in the expressions for both $I_{\rm stdy}$ and $I_{\rm trans}$, its 
presence in the steady-state contribution $I_{\rm stdy}$ does not prevent the
contribution to approach a steady-state value in the long-time limit. In
contrast, the value of $k_0$ is crucial for the transient contribution
$I_{\rm trans}$ to decay away. A zero $k_0$ would result in $I_{\rm trans}$ 
having a strong time-dependent zigzag-like behavior that would dominate the 
total energy current at all times. There would be no steady current flow 
even when $N\rightarrow\infty$. However, even a small on-site potential, 
say $k_0/k=0.1$, would result in the transient contribution to decay away 
in the long-time limit and leave only the contribution from the steady-state
term.


\end{document}